\documentclass[english,aps]{revtex4}
\usepackage[T1]{fontenc}
\usepackage[latin9]{inputenc}
\usepackage{babel}
\usepackage{amsmath}
\usepackage{graphicx}
\usepackage{esint}
\usepackage[unicode=true,pdfusetitle,
 bookmarks=true,bookmarksnumbered=false,bookmarksopen=false,
 breaklinks=false,pdfborder={0 0 1},backref=false,colorlinks=false]
 {hyperref}



\begin{document}

\title{Renormalization Group Improvement and Dynamical Breaking of Symmetry
in a Supersymmetric Chern-Simons-matter Model}

\author{A. G. Quinto}
\email{andres.quinto@ufabc.edu.br}

\author{A. F. Ferrari}
\email{alysson.ferrari@ufabc.edu.br}
\affiliation{\emph{Centro de Ciências Naturais e Humanas, Universidade Federal
do ABC - UFABC, Rua Santa Adélia, 166, 09210-170, Santo André, SP,
Brazil}}

\author{A. C. Lehum}
\email{lehum@ufpa.br}
\affiliation{\emph{Faculdade de Física, Universidade Federal do Pará, 66075-110,
Belém, Pará, Brazil}}

\begin{abstract}
In this work, we investigate the consequences of the Renormalization
Group Equation (RGE) in the determination of the effective superpotential
and the study of Dynamical Symmetry Breaking (DSB) in an $\mathcal{N}=1$
supersymmetric theory including an Abelian Chern-Simons superfield
coupled to $N$ scalar superfields in (2+1) dimensional spacetime.
The classical Lagrangian presents scale invariance, which is broken
by radiative corrections to the effective superpotential. We calculate
the effective superpotential up to two-loops by using the RGE and
the beta functions and anomalous dimensions known in the literature.
We then show how the RGE can be used to improve this calculation,
by summing up properly defined series of leading logs (LL), next-to-leading
logs (NLL) contributions, and so on... We conclude that even if the
RGE improvement procedure can indeed be applied in a supersymmetric
model, the effects of the consideration of the RGE are not so dramatic
as it happens in the non-supersymmetric case. 
\end{abstract}

\maketitle

\global\long\def\LL{{\rm LL}}

\global\long\def\NLL{{\rm NLL}}

\global\long\def\NNLL{{\rm N2LL}}

\section{Introduction}

Dynamical Symmetry Breaking (DSB) constitutes a very appealing scenario
in field theory, where quantum corrections are entirely responsible
for the appearance of nontrivial minima of the effective potential\,\cite{Coleman:1973jx}.
In the case of a classically scale invariant model, all mass scales
are generated by these quantum corrections and are fixed as functions
of the symmetry breaking scale. This scenario would be particularly
interesting in the Standard Model, but earlier calculations pointed
to a dead end: quantum corrections turned the effective potential
unstable, rendering DSB impossible\,\cite{Sher:1988mj}. However,
it has been shown that this conclusion, based on the effective potential
calculated up to the one-loop level, could be modified substantially
by using the Renormalization Group Equation (RGE)\,\cite{elias:2003zm,Chishtie:2005hr,PhysRevD.72.037902,Meissner:2006zh,Meissner:2007xv,Meissner:2008gj,Meissner:2008uw,AGQuinto}
to sum up infinite subsets of higher loop contributions to the effective
potential. More than a quantitative correction over the one-loop result,
this improvement lead to a new scenario, where DSB was operational\,\cite{elias:2003zm,Chishtie:2005hr}.
More recent calculations were able to include corrections up to five
loops in the effective potential\,\cite{Chishtie:2010ni,Steele:2012av},
bringing the prediction for the Higgs mass relatively close to the
experimental value indicated by the LHC (for other works regarding
conformal symmetry in the Standard Model see for example\,\cite{Englert:2013gz,Chun:2013soa}).

Besides being a viable ingredient to the Standard Model phenomenology,
DSB also occurs in other contexts, such as lower dimensional theories.
Particularly interesting are models involving the Chern-Simons (CS)
term in $\left(2+1\right)$ spacetime dimensions\,\cite{Deser:1981wh}.
The basic renormalization properties of such models have been studied
for quite some time\,\cite{Chen:1990sc,Avdeev:1991za,Chen:1992ee,ruizruiz:1996yf,Tan:1997ew,Alves:1999hw,dias:2003pw}.
We shall be particularly interested in models with scale invariance
at the classical level, that is, with a pure CS field coupled to massless
scalars and fermions, with Yukawa quartic interactions and scalar
$\varphi^{6}$ self-interactions. In these models, the one-loop corrections
calculated using the dimensional reduction scheme\,\cite{Siegel:1979wq}
are rather trivial, since no singularities appear, and no DSB happens
either; at the two-loop level, however, one finds renormalizable divergences.
Also, the two-loops effective potential $V_{eff}$ exhibits a nontrivial
minimum, signalizing the appearance of DSB. Due to the nontrivial
$\beta$ and $\gamma$ functions at two-loop level, one may obtain
an improvement in the calculation of $V_{eff}$ by imposing the RGE
\begin{align}
\left[\mu\frac{\partial}{\partial\mu}+\beta_{x}\frac{\partial}{\partial x}-\gamma_{\varphi}\phi\frac{\partial}{\partial\phi}\right]V_{eff}\left(\phi;\mu,\alpha_{i},L\right) & =0\,,\label{eq:RGE1}
\end{align}
where $x$ denotes collectively the coupling constants of the theory,
$\mu$ is the mass scale introduced by the regularization, $\gamma_{\varphi}$
is the anomalous dimension of scalar field $\varphi$, 
\begin{align}
L & =\ln\left[\frac{\phi^{2}}{\mu}\right],\label{eq:defL}
\end{align}
and $\phi$ is the vacuum expectation value of $\varphi$. This improved
effective potential was calculated in\,\cite{Dias:2010it}, and it
was shown to imply in considerable changes in the properties of DSB
in this model, thus providing another context where the consideration
of the RGE is essential to a proper analysis of the phase structure
of the model.

Our objective is to verify whether in supersymmetric models containing
the CS field, the consideration of the RGE also induces considerable
modifications in the scenario of DSB. Supersymmetric CS theories have
been studied for quite some time\,\cite{Avdeev:1991za,ruizruiz:1997jq,lehum:2007nf,Ferrari:2010ex,Lehum:2010tt},
and have recently attracted much attention due to their relation to
M2-branes\,\cite{Gaiotto:2007qi}. The superconformal field theory
describing multiple M2-branes is dual to the $D=11$ Supergravity
on $AdS_{4}\times S^{7}\sim[SO(2,3)/SO(1,3)]\times[SO(8)/SO(7)]\subset OSp(8|4)/[SO(1,3)\times SO(7)]$,
therefore the action for multiple M2-branes has $\mathcal{N}=8$ supersymmetry.
However, the on-shell degrees of freedom of this theory are exhausted
by bosons and physical fermions making its gauge sector to have no
on-shell degrees of freedom. These requirements are satisfied by a
Chern-Simons-matter theory called BLG theory\,\cite{Gustavsson:2008dy,Bagger:2007vi,Bagger:2007jr,Bandres:2008ry,Antonyan:2008jf},
which describes two M2-branes. Relaxing the requirement of manifest
$\mathcal{N}=8$ supersymmetry, this approach can be generalized to
a $\mathcal{N}=6$ Chern-Simons-matter theory with the gauge group
$U_{k}(N)\times U_{-k}(N)$ ($k$ and $-k$ are CS levels)\,\cite{Aharony:2008ug,Naghdi:2011ex},
which is expected to be enhanced to $\mathcal{N}=8$ for $k=1$ or
$k=2$\,\cite{Kwon:2009ar,Gustavsson:2009pm,Benna:2009xd}. The quantization
of such model was thoroughly studied in\,\cite{Faizal:2011en,Faizal:2014dca,Upadhyay:2014oda,Faizal:2012dj,Queiruga:2015fzn,Akerblom:2009gx,Bianchi:2009rf,Bianchi:2009ja,Bianchi:2010cx}.
Also, detailed calculations of the effective superpotential within
$\mathcal{N}=2$ superfield theories in three dimension have been
reported in\,\cite{Buchbinder:2012zd,Buchbinder:2015swa}. 

The first part of our work is the computation of the effective superpotential
of a generic supersymmetric CS theory coupled to matter superfields,
up to two-loops. To this end, we use the RGE and the $\beta$ and
$\gamma$ function calculated in\,\cite{Avdeev:1991za,Avdeev:1992jt},
thus avoiding the direct computation of any supergraph. With this
result in hand, we discuss how we can reorganize the expansion of
the effective superpotential in terms of Leading Logs (LL), Next-to-Leading
Logs (NLL) contributions, and so on, in a way that allows us to calculate
coefficients arising from higher orders corrections, thus improving
the two-loop evaluation of the effective superpotential. We are thus
able to find an improved effective superpotential, which will be used
to study DSB in our model. We will show that, contrary to what happens
in the non supersymmetric case\,\cite{Dias:2010it}, here the RGE
improvement leads only to slight modifications in the DSB scenario.

In this work, we shall focus on calculations done in the superfield
language\,\cite{gates:1983nr,buchbinder:1998qv}, in which supersymmetry
is manifest in all stages of the calculations. This paper is organized
as follows: in Sec.\,\ref{sec:Model}, we present our model and calculate
the effective superpotential with the knowledge of the renormalization
group functions found in the literature, together with the RGE. Section\,\ref{sec:Summing-up-Leading}
reviews the standard approach to RGE improvement of the effective
potential in four dimensional models, and section\,\ref{sec:RGE-Improvement-in3d}
is devoted to adapt this procedure to the supersymmetric three-dimensional
case. The resulting improved effective superpotential is used in Sec.\,\ref{sec:DYNAMICAL-BREAKING-OF}
to study DSB in our model. Section\,\ref{sec:Conclusions} presents
our conclusions and perspectives. Some explicit results and the Mathematica
code used to obtain them is given as a Supplementary Material to this
work.

\section{\label{sec:Model}Calculation of the Effective Superpotential}

Our starting point is the classical action in $\mathcal{N}=1$ superspace
of a Chern-Simons superfield $\Gamma_{\beta}$ coupled to $N$ massless
complex scalars superfields $\Phi_{a}$, with a quartic self-interaction,
\begin{eqnarray}
\mathcal{S} & = & \int d^{5}z\left\{ -\frac{1}{2}\Gamma^{\alpha}W_{\alpha}-\frac{1}{2}\overline{\nabla^{\alpha}\Phi_{a}}\nabla_{\alpha}\Phi_{a}+\frac{\lambda}{4}\left(\overline{\Phi_{a}}\Phi_{a}\right)^{2}\right\} ,\label{eq:M1}
\end{eqnarray}
where $W^{\alpha}=\frac{1}{2}D^{\beta}D^{\alpha}\Gamma_{\beta}$ is
the gauge superfield strength, $\nabla^{\alpha}=\left(D^{\alpha}-ig\,\Gamma^{\alpha}\right)$
is the gauge supercovariant derivative, and $a=1,\,...\,N$. We follow
the basic conventions for three-dimensional supersymmetry found in\,\cite{gates:1983nr}. 

The main object we shall be interested in studying is the three dimensional
effective superpotential\,\cite{Ferrari:2009zx}. To define this
object, we consider a shift in the $N$-th component of $\Phi_{a}$,
\begin{align}
\Phi_{N} & =\Phi_{N}^{q}+\sigma,\label{eq:M2}
\end{align}
by the background (constant) superfield $\sigma=\sigma_{1}-\theta^{2}\sigma_{2}$.
On general grounds, the effective superpotential can be cast as
\begin{equation}
\Gamma\left[\sigma\right]=\int d^{5}z\,K\left(\sigma\right)+\int d^{5}z\,F\left(\sigma,D_{\alpha}\sigma,D^{2}\sigma\right)\,.\label{eq:gammasplit}
\end{equation}
As discussed in\,\cite{Ferrari:2010ex}, the knowledge of $K$ is
sufficient for investigating the dynamical breaking of the gauge symmetry,
and consequential generation of a mass scale in the model, while the
study of a hypothetical supersymmetry breaking would involve also
the calculation of $F$\,\cite{Gallegos:2011ag,Gallegos:2011ux}.
For simplicity, in this work we will restrict ourselves on the calculation
of $K\left(\sigma\right)$, which we shall call the\emph{ effective
superpotential} from now on.

The effective superpotential $K\left(\sigma\right)$ is particularly
well suited for the approach we develop in this work, since we will
be able to calculate it by using a simple ansatz, using the information
already known from the literature regarding renormalization group
functions for the model\,(\ref{eq:M1}). The relevant results are
given in\,\cite{Avdeev:1991za}, from which we extract the two-loop
beta functions and anomalous dimension for the scalar superfield,\begin{subequations}\label{eq:RGEfunctions}
\begin{align}
\beta_{\lambda} & =c_{3}\lambda^{3}+c_{2}\lambda^{2}y+c_{1}\lambda y^{2}+c_{0}y^{3}\thinspace,\\
\beta_{y} & =0\thinspace,\\
\gamma_{\phi} & =d_{2}\lambda^{2}+d_{0}y^{2}\thinspace,
\end{align}
\end{subequations}in terms of the redefined gauge coupling constant
\begin{equation}
y=g^{2}\thinspace.
\end{equation}
The numerical coefficients present in\,(\ref{eq:RGEfunctions}) are
given by
\begin{multline}
c_{3}=\frac{3}{64\pi^{2}}\left(N+2\right),\thinspace c_{2}=\frac{1}{64\pi^{2}},\thinspace c_{1}=-\frac{2}{64\pi^{2}}\left(N+2\right)\thinspace,\\
c_{0}=-\frac{1}{64\pi^{2}}\left(N+3\right),\thinspace d_{2}=\frac{1}{4\left(64\pi^{2}\right)}\left(N+1\right),\thinspace d_{0}=-\frac{1}{4\left(64\pi^{2}\right)}\left(2N+3\right)\thinspace.
\end{multline}
These results are obtained from a two-loop computation of the divergent
vertex functions of the theory, since at one loop no divergences appear,
provided Feynman integrals are calculated by means of dimensional
regularization.

We shall use for $K\left(\sigma\right)$ the ansatz
\begin{equation}
K\left(\sigma\right)=-\frac{1}{4}\sigma^{4}S\left(L\right)\thinspace,\label{eq:KeffAnsatz1}
\end{equation}
where
\begin{equation}
S\left(L\right)=A\left(y,\lambda\right)+B\left(y,\lambda\right)L+C\left(y,\lambda\right)L^{2}+\cdots\thinspace,\label{eq:KeffAnsatz2}
\end{equation}
and $A,\thinspace B,\thinspace C,\thinspace\ldots$ are defined as
series in powers of the coupling constants $y$ and $\lambda$, and
$L$ is defined in\,\eqref{eq:defL}. We will eventually adopt a shorthand
notation where $x$ will denote any of the two couplings in our model,
so that a monomial like $y^{n}\lambda^{m}$ will be written as $x^{m+n}$.
Comparison with the tree level action\,(\ref{eq:M1}) shows us that
\begin{equation}
A\left(y,\lambda\right)=\lambda+{\cal O}\left(x^{2}\right)\thinspace,\label{eq:A}
\end{equation}
but actually the value of $A\left(y,\lambda\right)$ will be fixed
by the Coleman-Weinberg normalization of the effective superpotential,
\begin{equation}
\frac{1}{4!}\frac{d^{4}K\left(\sigma\right)}{d^{4}\sigma}=\frac{\lambda}{4}\thinspace,\label{eq:CWcondition}
\end{equation}
so only the $L$ dependent pieces of $K\left(\sigma\right)$, involving
$B,\thinspace C,\thinspace\ldots$, have to be calculated.

The other ingredient we will need is the RGE given in\,(\ref{eq:RGE1}).
From Eq,\,(\ref{eq:defL}) follows that $\partial_{L}=\frac{1}{2}\sigma\partial_{\sigma}=-\mu\partial_{\mu}$,
and inserting\,(\ref{eq:KeffAnsatz1}) into\,(\ref{eq:RGE1}), we
obtain an alternative form for the RGE,
\begin{equation}
\left[-\left(1+2\gamma_{\phi}\right)\partial_{L}+\beta_{\lambda}\partial_{\lambda}-4\gamma_{\phi}\right]S\left(L\right)=0\thinspace,\label{eq:RGE2}
\end{equation}
which will be used hereafter. 

Inserting the ansatz\,(\ref{eq:KeffAnsatz2}) in\,(\ref{eq:RGE2}),
and separating the resulting expression by orders of $L$, we obtain
a series of equations, of which we quote the first two:
\begin{equation}
-\left(1+2\gamma_{\phi}\right)B\left(y,\lambda\right)+\beta_{\lambda}\partial_{\lambda}A\left(y,\lambda\right)-4\gamma_{\phi}A\left(y,\lambda\right)=0\thinspace,\label{eq:orderL0}
\end{equation}
and
\begin{equation}
-2\left(1+2\gamma_{\phi}\right)C\left(y,\lambda\right)+\beta_{\lambda}\partial_{\lambda}B\left(y,\lambda\right)-4\gamma_{\phi}B\left(y,\lambda\right)=0\thinspace.\label{eq:orderL1}
\end{equation}
We now consider that all functions appearing in these equations are
defined as a series in powers of the couplings $x$, writing Eq.\,(\ref{eq:orderL0})
as
\begin{multline}
-\left(B^{\left(1\right)}+B^{\left(2\right)}+B^{\left(3\right)}+\cdots\right)-2\left(\gamma_{\phi}^{\left(2\right)}+\gamma_{\phi}^{\left(3\right)}+\cdots\right)\left(B^{\left(1\right)}+B^{\left(2\right)}+B^{\left(3\right)}+\cdots\right)\\
+\left(\beta_{\lambda}^{\left(3\right)}+\beta_{\lambda}^{\left(4\right)}+\cdots\right)\left(\partial_{\lambda}A^{\left(1\right)}+\partial_{\lambda}A^{\left(2\right)}+\cdots\right)\\
-4\left(\gamma_{\phi}^{\left(2\right)}+\gamma_{\phi}^{\left(3\right)}+\cdots\right)\left(A^{\left(1\right)}+A^{\left(2\right)}+\cdots\right)=0\thinspace,\label{eq:orderL0powers}
\end{multline}
where the numbers in the superscripts denote the power of $x$ of
each term. Since all terms of the previous equation start at order
$x^{3}$, except the first, we conclude that $B^{\left(1\right)}=B^{\left(2\right)}=0$,
and obtain the relation
\begin{equation}
B^{\left(3\right)}=\beta_{\lambda}^{\left(3\right)}-4\lambda\gamma_{\phi}^{\left(2\right)}\thinspace,
\end{equation}
after using Eq.\,(\ref{eq:A}). This last equation fixes the coefficients
of $B^{\left(3\right)}$ in terms of the (known) coefficients of $\beta_{\lambda}^{\left(3\right)}$
and $\gamma_{\phi}^{\left(2\right)}$, in the following form,
\begin{equation}
B^{\left(3\right)}=b_{3}\lambda^{3}+b_{2}\lambda^{2}y+b_{1}\lambda y^{2}+b_{0}y^{3}\thinspace,\label{eq:B3}
\end{equation}
where
\begin{equation}
b_{0}=c_{0};\thinspace b_{1}=c_{1}-4d_{0};\thinspace b_{2}=c_{2};\thinspace b_{3}=c_{3}-4d_{2}\thinspace.\label{eq:B3coefs}
\end{equation}
The corrections of the order $x^{3}L$ we have found for $S_{eff}$
could be obtained by a two-loop calculation of the effective superpotential,
using supergraph methods. Since we do not know the coefficients of
$\beta_{\lambda}^{\left(4\right)}$ and $\gamma_{\phi}^{\left(3\right)}$,
which would appear from higher loop corrections, we cannot use Eq.\,(\ref{eq:orderL0})
to calculate further coefficients of $B$ or $A$, so this equation
does not allow us to obtain information on higher-loops contributions
to $S_{eff}$. 

Now looking at Eq.\,(\ref{eq:orderL1}) expanded in power of the
couplings, one may conclude that $C\left(\lambda,y\right)$ starts
at order $x^{5}$, and obtain the relation,
\begin{equation}
C^{\left(5\right)}=\frac{1}{2}\beta^{\left(3\right)}\partial_{\lambda}B^{\left(3\right)}-2\gamma^{\left(2\right)}B^{\left(3\right)}\thinspace,\label{eq:C5}
\end{equation}
from which the coefficients of the form $x{}^{5}L^{2}$ of $S_{eff}$
are calculated from known coefficients of the beta functions, anomalous
dimension, and $B^{3}$. The end result is as follows,
\begin{multline}
C^{\left(5\right)}=\lambda^{5}\left(\frac{3}{2}c_{3}b_{3}-2d_{2}b_{3}\right)+\lambda^{4}y\left(c_{3}b_{2}+\frac{3}{2}c_{2}b_{3}-2d_{2}b_{2}\right)\\
+\lambda^{3}y^{2}\left(\frac{1}{2}c_{3}b_{1}+c_{2}b_{2}+\frac{3}{2}c_{1}b_{3}-2d_{0}b_{3}-2d_{2}b_{1}\right)\\
+\lambda^{2}y^{3}\left(\frac{1}{2}c_{2}b_{1}+c_{1}b_{2}+\frac{3}{2}c_{0}b_{3}-2d_{0}b_{2}-2d_{2}b_{0}\right)\\
+\lambda y^{4}\left(\frac{1}{2}c_{1}b_{1}+c_{0}b_{2}-2d_{0}b_{1}\right)+y^{5}\left(\frac{1}{2}c_{0}b_{1}-2d_{0}b_{0}\right)\thinspace.\label{eq:C5explicit}
\end{multline}
As a result, the RGE allows us to calculate terms of order $x^{5}L^{2}$
which, in our model, would appear only in a four loops explicit evaluation
of $V_{eff}$. Equation~(\ref{eq:orderL1}) does not provide us with
order $x^{6}L^{2},\thinspace x^{7}L^{2},\thinspace\ldots$ terms,
however, since we do not have knowledge of higher orders coefficients
of $\beta_{\lambda}$ and $\gamma_{\phi}$. 

At this point, it is clear that one could go on calculating order
$x^{7}L^{3},\thinspace x^{9}L^{4},\thinspace\ldots$ terms from Eqs.\,(\ref{eq:KeffAnsatz2})
and\,(\ref{eq:RGE2}), obtaining contributions to the effective superpotential
arising from higher loop orders, based just on the information we
have from the two loop calculation of $\beta_{\lambda}$ and $\gamma_{\phi}$.
In the next section, we will give an explanation of this pattern of
coefficients we are able to calculate, interpreting it as a leading
logs summation of the effective superpotential.

\section{\label{sec:Summing-up-Leading}RGE improvement and Dynamical Symmetry
Breaking: a short review of the four dimensional case}

We now review the procedure for the RGE improvement of the effective
potential calculation that was applied to the Standard Model in\,\cite{elias:2003zm,Chishtie:2005hr,Chishtie2011,Steele:2012av}
and to the non supersymmetric version of the model studied in this
work in\,\cite{Dias:2010it}. In doing so, further on we will be
able to pinpoint the differences we find in the supersymmetric three
dimensional case, still recognizing that the procedure outlined in
the previous section is essentially the same used in these works. 

Consider a scale invariant $\phi^{4}$ model in four spacetime dimensions,
coupled to other fermionic or gauge fields via a set of couplings
denoted collectively by $x$. The effective potential ${\cal V}_{eff}\left(\phi;\mu,x,{\cal L}\right)$
should satisfy the RGE
\begin{equation}
\left[\mu\frac{\partial}{\partial\mu}+\beta_{x}\frac{\partial}{\partial x}-\gamma_{\varphi}\phi\frac{\partial}{\partial\phi}\right]{\cal V}_{eff}\left(\phi;\mu,x,{\cal L}\right)=0\,,\label{eq:calRGE}
\end{equation}
where now 
\begin{equation}
{\cal L}=\ln\left[\frac{\phi^{2}}{\mu^{2}}\right]\thinspace.\label{eq:calLogCW}
\end{equation}
As before, we can rewrite this in a more convenient fashion by defining
\begin{equation}
{\cal V}_{eff}\left(\phi;\mu,x,{\cal L}\right)=\phi^{4}{\cal S}_{eff}\left(\mu,x,{\cal L}\right)\thinspace,\label{eq:defcalS}
\end{equation}
so that Eq.\,(\ref{eq:defcalS}) implies 
\begin{equation}
\left[-\left(2+2\gamma_{\phi}\right)\frac{\partial}{\partial{\cal L}}+\beta_{x}\frac{\partial}{\partial x}-4\gamma_{\varphi}\right]{\cal S}_{eff}\left(\mu,x,{\cal L}\right)=0\,.\label{eq:calRGE2}
\end{equation}

The central point of the general approach to RGE improvement discussed
in the aforementioned references is to reorganize the contributions
to ${\cal S}_{eff}\left(\mu,x,{\cal L}\right)$ arising from different
loop orders according to the difference between the aggregate power
of the couplings $x$ and the logs ${\cal L}$, that is to say,
\begin{equation}
{\cal S}_{eff}\left(x,{\cal L}\right)={\cal S}_{eff}^{\LL}\left(x,{\cal L}\right)+{\cal S}_{eff}^{\NLL}\left(x,{\cal L}\right)+\cdots\,,\label{eq:calSLL}
\end{equation}
where ${\cal S}_{eff}^{\LL}$ contains the leading logs contributions,
\begin{equation}
{\cal S}_{eff}^{\LL}\left(x,{\cal L}\right)=\sum_{n\geq1}{\cal C}_{n}^{\LL}x^{n}{\cal L}^{n-1}\,,\label{eq:defSLL}
\end{equation}
${\cal S}_{eff}^{\NLL}$ contains the next to leading logs terms,
\begin{equation}
{\cal S}_{eff}^{\NLL}\left(x,{\cal L}\right)=\sum_{n\geq2}{\cal C}_{n}^{\NLL}x^{n}{\cal L}^{n-2}\,,\label{eq:defSNLL}
\end{equation}
and so on. Insertion of the ansatz\,(\ref{eq:calSLL}) in the RGE\,(\ref{eq:calRGE2})
gives a set of coupled differential equations, of which we quote the
first two,
\begin{equation}
\left[-2\frac{\partial}{\partial{\cal L}}+\beta_{x}^{\left(2\right)}\frac{\partial}{\partial x}\right]{\cal S}_{eff}^{\LL}\left(x,{\cal L}\right)=0\,,\label{eq:eqs1}
\end{equation}
and
\begin{equation}
\left[-2\frac{\partial}{\partial{\cal L}}+\beta_{x}^{\left(2\right)}\frac{\partial}{\partial x}\right]{\cal S}_{eff}^{\NLL}\left(x,{\cal L}\right)+\left[\beta_{x}^{\left(3\right)}\frac{\partial}{\partial x}-4\gamma_{\varphi}^{\left(2\right)}\right]{\cal S}_{eff}^{\LL}\left(x,{\cal L}\right)=0\,.\label{eq:eqs2}
\end{equation}
Equation\,(\ref{eq:eqs1}) results in a first order difference equation
for ${\cal C}_{n}^{\LL}$, so the knowledge of the initial coefficient
${\cal C}_{1}^{\LL}$ and the order $x^{2}$ contribution to the beta
function from loop calculations, allows one to calculate all ${\cal C}_{n}^{\LL}$,
therefore summing up all the leading logs contributions to the effective
potential. This summation was the key to making the DSB scenario viable
in the scale invariant Standard Model as shown in\,\cite{elias:2003zm}.
One does not need to stop at this point, however, since Eq.\,(\ref{eq:eqs2})
can also be used to sum up the next to leading logs, after ${\cal S}_{eff}^{\LL}$
was calculated, provided one knows the first coefficient ${\cal C}_{2}^{\NLL}$
of the series, as well as $\beta_{x}^{\left(3\right)}$ and $\gamma_{\varphi}^{\left(2\right)}$.
That means one can sum up sequentially several subseries of coefficients
contributing to the effective potential, until exhausting the perturbative
information encoded in $\beta_{x}$, $\gamma_{\phi}$, and the $V_{eff}$
calculated up to a certain loop order. This is a systematical procedure
to extract the maximum amount of information concerning the effective
potential from a perturbative calculation. 

One important technical detail is that the renormalization group functions
are usually calculated in the Minimal Subtraction (MS) renormalization
scheme, and they need to be adapted to the procedure outlined in this
section, as it was first pointed out in\,\cite{Ford:1991hw}. For
simplicity, let us consider the case of a theory with a single coupling
$x$. In the MS scheme, divergent integrals (in four spacetime dimension)
appear with a factor
\begin{equation}
\tilde{{\cal L}}=\ln\left[\frac{x\,\phi^{2}}{2\,\mu^{2}}\right]\thinspace,\label{eq:calLogMS}
\end{equation}
while in the so-called Coleman-Weinberg (CW) scheme, the effective
potential depends on a log of the form\,(\ref{eq:calLogCW}). Both
schemes can be related by a redefinition of the mass scale $\mu$,
\begin{equation}
\mu_{MS}^{2}=f\left(x\right)\mu_{CW}^{2}\thinspace,
\end{equation}
which can be shown to imply in the following relation between the
beta functions in both schemes,
\begin{alignat}{1}
\beta_{CW} & =\beta_{MS}\left(1-\frac{1}{2}\beta_{MS}\,\partial_{x}\ln f\right)^{-1}\thinspace.\label{eq:relationbetas}
\end{alignat}
In four spacetime dimensions, divergences usually start at one loop,
generating order $x^{2}$ contributions to $\beta_{MS}$, therefore,
\begin{align}
\beta_{CW} & =\left(\beta_{MS}^{\left(2\right)}+\beta_{MS}^{\left(3\right)}+\cdots\right)\left(1+{\cal O}\left(x\right)\right)\nonumber \\
 & =\beta_{MS}^{\left(2\right)}+{\cal O}\left(x^{3}\right)\thinspace.\label{eq:relCWMS}
\end{align}
The conclusion is that at one loop level, both beta functions can
be used interchangeably, but if calculations are done at two loops
or more, one has to adapt the MS functions to be used in the calculation
of the CW effective potential. The same reasoning concerning the beta
functions can be applied to the anomalous dimension, with similar
conclusions. 

To gain further insight into this problem, we present the following
argument: in the MS and CW schemes, the effective potential would
be calculated at one loop level in the forms
\begin{equation}
V_{MS}=\phi^{4}\left(\tilde{A}\left(x\right)+\tilde{B}\left(x\right)\tilde{{\cal L}}\right)\thinspace,\label{eq:VMS1}
\end{equation}
and
\begin{equation}
V_{CW}=\phi^{4}\left(A\left(x\right)+B\left(x\right){\cal L}\right)\thinspace.\label{eq:VCW1}
\end{equation}
From Eqs.\,(\ref{eq:calLogCW}) and\,(\ref{eq:calLogMS}), we have
\begin{equation}
\tilde{{\cal L}}={\cal L}+\ln\left[\frac{x}{2}\right]\thinspace,\label{eq:relLogs}
\end{equation}
and therefore one can rewrite $V_{MS}$ in a form compatible with
the CW scheme as follows,
\begin{equation}
V_{MS}=\phi^{4}\left[\left(\tilde{A}\left(x\right)+\ln\frac{x}{2}\tilde{B}\left(x\right)\right)+\tilde{B}\left(x\right){\cal L}\right]\thinspace.
\end{equation}
Since the value of $A\left(x\right)$ is immaterial in the CW potential,
being fixed by the CW condition\,(\ref{eq:CWcondition}), we conclude
that $\tilde{B}\left(x\right)=B\left(x\right)$ and that both $V_{MS}$
and $V_{CW}$ end up giving identical results at one loop. At two
loops, however, $V_{MS}$ contains a term of the form $\tilde{C}\left(x\right)\tilde{{\cal L}}^{2}$,
so after employing\,(\ref{eq:relLogs}), one would find a difference
in the relevant term proportional to ${\cal L}$, meaning both potentials
are not equivalent at two loops. The net result is that, at the two
loop level, the RGE can be used to relate renormalization group functions
and the effective potential in the CW and the MS scheme, but not interchangeably.

\section{\label{sec:RGE-Improvement-in3d}RGE Improvement in the Three Dimensional
Supersymmetric Case}

Now we discuss how to adapt the procedure outlined in section\,\ref{sec:Summing-up-Leading}
to our model. First of all, we consider the problem of interchangeability
of MS and CW renormalization group functions when using the RGE to
calculate the effective potential. In the supersymmetric three-dimensional
model considered by us, divergences only start at two loops, and the
beta functions start at order $x^{3}$. This means that instead of
Eq.\,\ref{eq:relCWMS} we have
\begin{align}
\beta_{CW} & =\left(\beta_{MS}^{\left(3\right)}+\beta_{MS}^{\left(4\right)}+\cdots\right)\left(1+{\cal O}\left(x\right)\right)\nonumber \\
 & =\beta_{MS}^{\left(3\right)}+{\cal O}\left(x^{4}\right)\thinspace.
\end{align}
Also, $V_{eff}$ acquires a term proportional to $\tilde{{\cal L}}^{2}$
only at loop orders greater than two. This means we are safe to use
interchangeably functions calculated in the MS and the CW scheme,
as we have done in section\,\ref{sec:Model}.

It is still not clear that the series of terms we calculated in the
previous section, of orders $x^{2n+1}L^{n}$, have any relation to
the leading logs summation described in section\,\ref{sec:Summing-up-Leading}.
Indeed, by repeating the steps outlined in the start of that section
for our model, the fact that $\beta_{\lambda}^{\left(2\right)}=0$
together with Eq.\,(\ref{eq:eqs1}) would imply ${\cal C}_{n}^{LL}=0$
for $n>1$, which in its turn would also trivialize Eq.\,(\ref{eq:eqs2}).
The conclusion would be that the RGE does not allow us to calculate
any new contribution for the effective superpotential.

Actually, this apparent problem is a consequence of the particular
pattern of divergences that appear in our model, whenever we use dimensional
regularization to evaluate Feynman integrals. In four dimensional
non supersymmetric theories, divergences in general occur at any loop
order $n$, the leading logs being of the order $x^{n+1}L^{n}$. In
three dimensional supersymmetric models, divergences start only at
two loops, and are of the order $x^{3}L$. At three loops, the only
divergences arise from two loops subdiagrams, of the order $x^{4}L$.
At four loops, we find again superficial divergent diagrams, of order
$x^{5}L^{2}$, while five loops diagrams contain at the most four
and two loops divergent subdiagrams, of order $x^{6}L^{2}$ and $x^{6}L$.
This pattern suggests that new superficial divergences appear only
at even loops, and are of the order $x^{2n+1}L^{n}$, and these terms
should be identified as ``leading logs'' in our case, despite the
fact that the difference between the power of coupling constants and
logs is not the same for all terms. Careful consideration of this
divergence pattern suggests for supersymmetric three dimensional models
the definition,
\begin{equation}
S_{eff}\left(x,L\right)=S_{eff}^{\LL}\left(x,L\right)+S_{eff}^{\NLL}\left(x,L\right)+\cdots\,,\label{eq:ansatz3D}
\end{equation}
where leading logs contributions are of the form
\begin{equation}
S_{eff}^{\LL}\left(x,L\right)=\sum_{n\geq0}C_{n}^{\NLL}x^{2n+1}L^{n}\thinspace,\label{eq:LL3d}
\end{equation}
next to leading logs are given by
\begin{equation}
S_{eff}^{\NLL}\left(x,L\right)=\sum_{n\geq0}\left(C_{n}^{\NLL}x^{2n+2}L^{n}+D_{n}^{\NLL}x^{2n+3}L^{n}\right)\thinspace,\label{eq:NLL3d}
\end{equation}
and so on. Inserting this ansatz into the RGE\,(\ref{eq:RGE2}) gives
us
\begin{multline}
\sum_{n\geq0}\left[\left(-\left(n+1\right)C_{n+1}^{\LL}+\left(2n+1\right)\beta^{3}C_{n}^{\LL}-4\gamma_{\phi}^{\left(2\right)}C_{n}^{\LL}\right)x^{2n+3}L^{n}\right.\\
\left.+{\cal O}\left(x^{2n+5}L^{n}\right)\right]=0\thinspace,\label{eq:eqs3}
\end{multline}
which, very much like Eq.\,(\ref{eq:eqs1}), provides a first order
difference equation for $C_{n}^{\LL}$, now involving the order $x^{3}$
terms in the beta function, as well as the order $x^{2}$ terms of
the anomalous dimension. From this equation, the whole series of leading
logs terms may be (in principle) summed up, and $S_{eff}^{\LL}\left(x,L\right)$
determined from the two loop information we have at hand. Looking
at other coefficients of the sum in Eq.\,(\ref{eq:eqs3}) would provide
equations for the calculation of next to leading log contributions,
and so on. The result is that the leading logs summation procedure
can be applied to three dimensional supersymmetric models, yet with
nontrivial modifications, taking into account the peculiar divergence
structure of such models.

To actually apply this technique to our model, one has to generalize
the equations in last paragraph to the case of two couplings, which
involves dealing with double sums of the form
\begin{equation}
S_{eff}^{\LL}\left(\lambda,y,L\right)=\sum_{n\geq0}\sum_{\,0\leq\ell\leq2n+1}C_{n,\,2n+1-\ell}^{\LL}\,\lambda^{\ell}\,y^{2n+1-\ell}L^{n}\thinspace.
\end{equation}
We do not quote here the algebraic details, but we developed a Mathematica
code to calculate the coefficients $C^{\LL}$ up to an arbitrary (finite)
order. With this code we could reproduce, as a consistency check,
the result given in Eq.\,(\ref{eq:C5explicit}), as well as calculating
corrections to $S_{eff}^{\LL}$ up to the order $x^{41}L^{20}$ in
a few seconds. This result will be used, in the next section, to study
the modifications introduced by the leading logs summation in the
DSB in our model. The code as well as the explicit results are available
as a Supplementary Material to this paper.

\section{\label{sec:DYNAMICAL-BREAKING-OF}Dynamical Breaking of Symmetry }

In this Section we study the dynamical breaking of the conformal symmetry
that occurs in the present theory, based on the improved effective
superpotential that was obtained in the previous sections, by summing
up leading logs contributions up to the order $x^{41}L^{20}$. More
explicitly, we consider,
\begin{align}
K_{eff}^{I}\left(\sigma\right) & =-\frac{1}{4}\sigma^{4}\left[S_{eff}^{\LL}\left(\lambda,y,L\right)+\rho\right],\label{eq:KalherianInprove}
\end{align}
$\rho$ being a finite renormalization constant. The constant $\rho$
is fixed using the CW normalization condition\,(\ref{eq:CWcondition}).
Requiring that the $K_{eff}^{I}\left(\sigma\right)$ has a minimum
at $\sigma^{2}=\mu$ means that
\begin{align}
\left.\frac{d}{d\sigma}K_{eff}^{I}\left(\sigma\right)\right|_{\sigma^{2}=\mu} & =0\thinspace,\label{eq:minimo}
\end{align}
which can be used to determine the value of $\lambda$ as a function
of the free parameters $y$ and $N$. 

\begin{figure}
\begin{centering}
\includegraphics{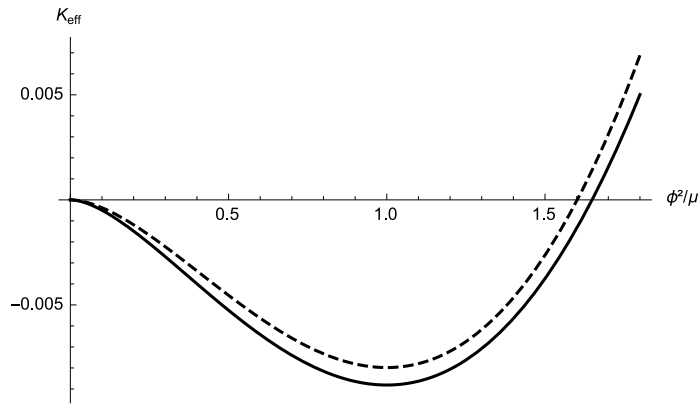}
\par\end{centering}

\caption{\label{fig:UnimproveandImprove}Comparison of the unimproved (solid
line) and improved (dashed line) effective superpotential, for $y=0.8$
and $N=90$. }
\end{figure}

Upon explicit calculation, Eq.\,(\ref{eq:minimo}) turns out to be
a polynomial equation in $\lambda$, and among its solutions we look
for those which are real and positive, and correspond to a minimum
of the potential, i.e., 
\begin{align}
M_{\Sigma}=\left.\frac{d^{2}}{d\sigma^{2}}K_{eff}^{I}\left(\sigma\right)\right|_{\sigma^{2}=\mu} & >0\thinspace.
\end{align}
This procedure was implemented in a \emph{Mathematica} program, and
we verified that it can be performed for any value of $y<1$ and $N$.
That means DSB is operational in this model for any reasonable value
of its parameters. As an example: choosing $y=0.1$ and $N=1$, we
found that $\lambda^{I}=0.000023224294742$. To compare, by choosing
the same values of $y$ and $N$, but using the unimproved two-loop
effective superpotential, including only corrections up to order $x^{3}L$,
we find $\lambda=0.00002322078553849$. The difference between the
two values being only of order $0.015\%$, we say that the improvement
of the effective superpotential by means of the summation of the leading
logs contributions provides only a small quantitative change on the
parameters of the DSB. This is rather different from the scenario
found in four dimensional models, or even the non supersymmetric version
of the same model considered by us, where the RGE improvement provided
substantial qualitative changes in the phase structure of the DSB\,\cite{Dias:2010it}.
The incremental aspect of the improvement, in the present case, can
also be seen by plotting both the improved and unimproved effective
superpotentials as in Fig.\,\ref{fig:UnimproveandImprove}, where
only by choosing relatively high values of $y$ and $N$ we were able
to get two graphs that do not superimpose. 

\section{\label{sec:Conclusions}Conclusions}

The mechanism of symmetry breaking is central for the formulation
of a consistent quantum field theory of the known elementary interactions,
and the possibility that quantum corrections of a symmetric potential
could alone induce such symmetry breaking is a rather interesting
one, not only for its mathematical elegance, but also for physical
reasons. Recently, for example, a mechanism of dynamical symmetry
breaking in a scale-invariant version of the Standard Model is being
discussed as a viable mechanism for generating a mass for the Higgs
particle compatible with experimental observations. The idea of using
the RGE to improve the calculation of the effective potential, summing
up terms arising from higher loop orders organized as leading logarithms,
next to leading logarithms, and so on, is central to this approach.
We have shown how this program can be applied to a supersymmetric
model in the superfield formalism, which is the main technical result
of this paper. 

We discussed an ${\cal N}=1$ supersymmetric Abelian Chern-Simons
model coupled to an arbitrary number of scalar and fermion superfields.
Matter fields are assumed to be minimally coupled to the CS field,
together with quartic self-interaction. The use of the renormalization
group functions calculated in\,\cite{Avdeev:1991za} together with
the RGE allowed a calculation of the improved effective superpotential,
that can be used to study DSB in our model. The end result was that
DSB is operational for all reasonable values of the free parameters,
and that the RGE improvement produces only a small quantitative change
in the properties of the model. 

In this particular model, therefore, the effects of the RGE improvement
were not so dramatic as in its non supersymmetric counterpart, however
the question remains whether the same might happen in different models.
It begs to say, however, that we do not expect this technique to be
directly applicable to four dimensional supersymmetric models, for
which non renormalization theorems in general forbid DSB.

One final remark is in order. In this work, we used the superfield
formalism for the evaluation of the effective superpotential and the
study of the phase structure of the model in a manifestly supersymmetric
way. One might wonder about the effective scalar potential $V_{eff}$,
i.e., the effective potential of the scalar component of the constant
background superfield $\sigma$. $V_{eff}$ should be calculated from
the full effective superpotential $\Gamma\left[\sigma\right]$ as
described in\,\cite{Ferrari:2010ex}. In this case, one should be
careful in isolating the contribution from the auxiliary field effective
superpotential $F$ (see Eq.\,(\ref{eq:gammasplit})), and also use
the beta functions appropriate for the component fields (the superfield
quartic coupling $\frac{\lambda}{4}\left(\overline{\Phi_{a}}\Phi_{a}\right)^{2}$
translates into a coupling $\frac{\lambda^{2}}{4}\left(\overline{\varphi_{a}}\varphi_{a}\right)^{3}$
in the component formalism, for example) for the RGE improvement.
This approach would be natural if one were to consider an important
aspect that was left out of this paper because of its technical complexity:
the inclusion\textbackslash{} of the effects of the auxiliary field
effective superpotential $F$, which would allow us to investigate
the possibility of spontaneous breaking of supersymmetry. This would
deserve a separate investigation. Since in the approximation we are
considering there is no possibility of supersymmetry breaking, it
is simpler to consider the $K\left(\sigma\right)$ effective superpotential
as the central object of our study, as it was done in our work.

\bigskip{}

\textbf{Acknowledgments. }The authors would like to thank Professor
Alex Gomes for useful discussions. This work was supported by Conselho
Nacional de Desenvolvimento Científico e Tecnológico (CNPq), Fundação
de Amparo a Pesquisa do Estado de São Paulo (FAPESP), Coordenação
de Aperfeiçoamento de Pessoal de Nível Superior (CAPES), and Fundação
de Apoio à Pesquisa do Rio Grande do Norte (FAPERN), via the following
grants: CNPq 482874/2013-9, FAPESP 2013/22079-8 and 2014/24672-0 (AFF),
CAPES PhD grant (AGQ), CNPq 308218/2013-2 and FAPERN (ACL).

\end{document}